\documentclass[12pt,one column,showpacs,preprintnumbers,pra,aps]{revtex4}
\usepackage{amsmath}
\usepackage{graphicx}
\usepackage{dcolumn}
\usepackage{array}
\usepackage{bm}

\begin{document}
\title{On the quantum master equation for Bogoliubov-BCS quasiparticles}
\author{ Chun-Feng Huang${^{(1),(2),(3)}}$ and Keh-Ning Huang${^{(1),(4)}}$}

\affiliation{$^{(1)}$ Department of Physics, National Taiwan University, Taipei, Taiwan, R. O. C.}
\affiliation{$^{(2)}$ National Measurement Laboratory, Center for Measurement Standards, Industrial Technology Research Institute, Hsinchu, Taiwan, R. O. C. }
\affiliation{$^{(3)}$ 2nd Patent Division, Intellectual Property Office, MOEA}
\affiliation{$^{(4)}$ Institute of Atomic and Molecular Sciences, Academia Sinica, Taipei, Taiwan, R. O. C.}
\date{\today}
\begin{abstract}
The quantum master equation is introduced for the density matrix representing Bogoliubov-BCS quasiparticles. A constraint to relate the loss and gain factors is taken into account to preserve the form of the density matrix. Such an equation can be reduced to the semiclassical equation, and can be extended for the coexistence of different order parameters.
\end{abstract}
\pacs{34.80.Dp}
\maketitle
\section{Introduction}
Considerable efforts have been made to develop quantum kinetic approaches. \cite{Ohtsuki,Louisell,Kampen,Lindblad,Alicki,Huang1,Gebauer1,Huang2,Pershin,Yau,Jaynes,Swenson,Gough,Michael2,Smirnov,Benatti,Huang3} Although semiclassical kinetic approaches can be used to model many irreversible processes, quantum corrections should be taken into account when we consider the systems of the nano-scale. Different master equations have been introduced for such corrections by including quantum relaxation terms in addition to the Liouville term. \cite{Ohtsuki,Louisell,Kampen,Lindblad,Alicki,Huang1,Gebauer1,Huang2,Pershin,Yau} The Markoff master equation is successful in quantum optics, \cite{Ohtsuki,Louisell} and the equation of Lindblad form \cite{Kampen,Lindblad} is derived by considering suitable assumptions. It is also discussed in the literature how to construct the quantum master equations to include Fermi and Bose properties. \cite{Alicki,Huang1} The nonlinear relaxation terms have been introduced in Refs. \cite{Huang1} and \cite{Gebauer1} for non-interacting identical fermions, and we can see the equivalence after some calculations. In addition to the formal derivation, we can obtain such nonlinear terms intuitively by considering both the lifetimes of particles and holes based on the conservation of the number of particles in each transition. \cite{Huang1,Huang2} Here holes are vacancies of any orbitals. The lifetimes of particles and holes describe the loss and gain of particles, respectively, and the equations introduced in Refs. \cite{Alicki,Huang1,Gebauer1} for fermions are symmetric with respect to particles and holes. 

Because the nonlinear relaxation terms in Refs. \cite{Huang1} and \cite{Gebauer1} are for non-interacting fermions, it is natural that they can become invalid when many-body effects are important. \cite{Pershin} Such effects have been taken into account under significant two-body interactions, which may induce the formation of Bogoliubov-BCS quasiparticles \cite{Barankov,Nazario,Ohta,Bogoliubov,Valatin,Bena,Laughlin}. Such fermionic quasiparticles have been successfully introduced to understand superconductors. While the number of particles is conserved in the conventional noninteracting models, the conservation is not necessary for quasiparticles. Semiclassical master equations have been proposed to model Bogoliubov-BCS quasiparticles when the wave properties can be neglected. \cite{Aronov,Entin-Wohlman} On the other hand, WKB approximation \cite{Aronov,WKB} based on Bogoliubov-de Gennes equation \cite{Galperin} provides an analytic way to understand the wave behaviors. \cite{Bardeen1} To describe the quantum wave properties when the relaxation effects are incorporated, it is of the fundamental interest to extend the quantum master equations to Bogoliubov-BCS quasiparticles. \cite{Huang2,Dubi2}  

In this paper, the quantum master equation is extended for the fermionic density matrix representing Bogoliubov-BCS quasiparticles. Without the loss of generality, we focus on the quasiparticles due to the pairings of superconducting electrons in the coordinate space. In section II, we discuss how to construct the loss and gain factors for such a density matrix based on the general form introduced in Ref. \cite{Huang1}. Different types of order parameters \cite{Laughlin} can be taken into account in the extended equation, as shown in section III. The discussion and conclusion are made in section IV and V, respectively.

\section{A quantum master equation for Bogoliubov-BCS quasiparticles}
The master equation of the following form
\begin{eqnarray} 
\frac{\partial}{\partial t} \hat{\rho} (t) = i [ \hat{\rho} (t), \hat{H} (t) ] - \{ \hat{\rho} (t) , \hat{\Gamma} (t) \} + \{ \hat{I} - \hat{\rho} (t) , \hat{\Gamma} ^{\prime }(t) \}
\end{eqnarray}
has been introduced in Ref. \cite{Huang1} for identical fermions. Here $\hat{\rho} (t)$ and $\hat{H} (t)$ represent the density matrix and Hamiltonian for the considered fermions, $\hat{I}$ is the identity operator, and we denote $[\hat{O} _{1}, \hat{O} _{2}]$ and $\{ \hat{O} _{1}, \hat{O} _{2} \}$ as the commutator and anticommutator for any two operators $\hat{O} _{1}$ and $\hat{O} _{2}$. In this paper, we take the reduced Planck constant $\hbar=1$. We require that $\hat{\Gamma} (t)$ and $\hat{\Gamma} ^{ \prime } (t)$ are both positive self-adjoint operators such that the above equation can preserve the positivity and Pauli's exclusion principle under suitable assumptions. \cite{Huang2} The operator $\hat{I}  - \hat{\rho} (t)$ represents the density matrix for the corresponding holes, and the last two terms in Eq. (1) describe the lifetimes of particles and holes for the irreversible effects due to the relaxation and/or excitation. \cite{Huang1} To extend Eq. (1) as that \cite{Huang1,Alicki} for bosons, we just need to replace the last term by $\{ \hat{I} + \hat{\rho} (t) , \hat{\Gamma} ^{ \prime }(t) \}$. We can see from Refs. \cite{Alicki,Huang1,Gebauer1} that the form of Eq. (1) can be derived by different ways.

To see the meanings of the last two terms in Eq. (1) explicitly, consider the noninteracting spin-unresolved electrons in a single band of one finite cube with the volume $V$ under the periodic boundary condition. In addition, assume that the Hamiltonian (defined in the many-body space) can be approximated as the time-independent operator 
\begin{eqnarray}
H _{o} = \sum _{ {\bf k}  , \sigma} \varepsilon _{{\bf k}} c ^{\dagger} _{{\bf k} , \sigma} c _{ {\bf k} , \sigma }
\end{eqnarray} 
around a specific time $t_{1}$. Here $\sigma$ is the spin orientation, the wavevector $ {\bf k} $ is quantized in the Brillouin zone because of the boundary condition, $c _{ {\bf k} , \uparrow }$ ($c _{ {\bf k} , \downarrow }$) represents the annihilator for the spin-up (spin-down) electron with the plane wave $u _{e} ( {\bf r} ; {\bf k} ) \equiv \frac{1}{ V ^{1/2} } exp( i {\bf k} \cdot {\bf r}  )$ as the spatial wavefunction, and the real number $\varepsilon _{{\bf k}}$ represents the eigenenergy of the spin-unresolved orbital corresponding to $c _{{\bf k} , \sigma}$.  Under Eq. (2), in the coordinate space the Hamiltonian $\hat{H} _{e} (t)$ for electrons is
\begin{eqnarray}
\hat{H} _{o} =  \sum _{ {\bf k}  , \sigma} \varepsilon _{{\bf k}} u _{e} ( {\bf r} ; {\bf k} ) u _{e} ^{\ast} ( {\bf r} ^{\prime}  ; {\bf k} ),
\end{eqnarray}
as $t \sim t _{1}$. (Taking  $| {\bf k} \sigma \rangle$ as the ket with $ u _{e} ( {\bf r} ; {\bf k} ) $ as the spatial wavefunction,  $\hat{H} _{o} =  \sum _{ {\bf k}  , \sigma} \varepsilon _{{\bf k}} | {\bf k} \sigma \rangle \langle {\bf k} \sigma  | $.) Because of the unresolved spin-splitting, $\langle c ^{\dagger} _{{\bf k} , \sigma} c _{{\bf k} ^{\prime} , \sigma ^{\prime} } \rangle = \langle c ^{\dagger} _{{\bf k} , -\sigma} c _{{\bf k} ^{\prime} , -\sigma ^{\prime} } \rangle \delta _{\sigma , \sigma ^{\prime} }$ and we do not need to consider the spin-orientation. Here we take $\langle A \rangle$ as the expectation value of any (many-body) operator $A$ with respect to the total (many-body) density matrix, which describes both the reservoir and the considered system. When the transitions are between the eigenorbitals of $\hat{H} _{o}$, we shall replace the operators $\hat{\Gamma} (t)$ and $\hat{\Gamma} ^{ \prime } (t)$ in Eq. (1) as   
\begin{eqnarray}
\hat{\Gamma} _{e} (t) = \sum _{{\bf k} } \alpha _{ e } ({\bf k};t)  u_{ e } ({\bf r} ;{\bf k}) u_{ e } ^{\ast} ({\bf r}^{\prime} ;{\bf k})
\text{ and }
\hat{\Gamma} ^{ \prime } _{e} (t) = \sum _{{\bf k} } \beta _{ e} ({\bf k};t) u_{e} ({\bf r};{\bf k}) u_{e} ^{\ast} ({\bf r} ^{\prime};{\bf k})
\end{eqnarray}
in the coordinate space when $t \sim t _{1}$. Here $\alpha _{ e } ({\bf k};t)$ and $ \beta _{ e } ({\bf k};t)$ are the positive coefficients for the loss and gain of electrons in orbital ${\bf k}$, and for any complex number $z$ we denote $z ^{\ast}$ as its complex conjugate in this paper. Let $F _{e} ( {\bf k};t) = \langle c ^{\dagger} _{{\bf k} \sigma} c _{{\bf k} \sigma } \rangle $ as the number of electrons annihilated by $c _{ {\bf k} , \sigma }$ at time $t$. We have from Eqs. (1) and (4)
\begin{eqnarray}
\frac{\partial}{\partial t} F _{e} ( {\bf k};t) = - 2 \alpha _{ e } ({\bf k};t) F _{e} ( {\bf k} ;t) + 2 \beta _{ e } ({\bf k};t) ( 1 - F _{e} ( {\bf k} ;t) ).
\end{eqnarray}
The function $F _{e} ( {\bf k} ;t)$ follows $0 \leq F _{e}( {\bf k};t) \leq 1$ because electrons are fermions, and the first and second terms at the right hand side of the above equation is to decrease and increase the occupation number in orbital ${\bf k}$. In Eq. (1), therefore, the last two terms are responsible for  the loss and gain. The above equation is just the nonlinear semiclassical master equation \cite{Liboff} if 
\begin{eqnarray}
\alpha _{ e } ({\bf k};t)= \sum _{{\bf k} ^{\prime}  } \frac{ \omega _{ {\bf k}^{\prime} {\bf k} } }{2} ( 1 - F _{e} ({\bf k} ^{\prime};  t ) ) \text{ and } 
\beta _{ e } ({\bf k};t)= \sum _{ {\bf k}^{\prime}  } \frac{ \omega _{ {\bf k} {\bf k} ^{\prime} } }{2} F _{e} ( {\bf k} ^{\prime};t ).
\end{eqnarray}
Here $ \omega _{ {\bf k} {\bf k}^{\prime} } $ denotes the positive coefficient for the transition from ${\bf k} ^{\prime}$ to $ {\bf k}$ if ${\bf k} \not= {\bf k} ^{\prime}$, and we take $\omega _{ {\bf k} {\bf k} }=0$ for all ${\bf k}$. Because the semiclssical master equation is not linear, it is natural that the last two terms in Eq. (1) is nonlinear. \cite{Huang1,Gebauer1}

To extend Eq. (1) for Bogoliubov-BCS quasiparticles in superconudctors, we note that the wavefunction 
\begin{eqnarray}
{\bf f} _{b}=\left( \begin{array}{c} f _{1} \\ f _{2} \end{array} \right)
\end{eqnarray}
for such quasiparticles is the direct sum of the electron- and hole-components defined in the coordinate space. For convenience, we denote ${\bf f} _{b} ^{\dagger} \equiv (f _{1} ^{\ast} \text{ , } f ^{\ast} _{2})$. It is known that the effective Hamiltonian is of the form \cite{Galperin,Valatin}
\begin{eqnarray}
\hat{H} _{b} (t)=\left[
\begin{array}{c}
\hat{ {\cal H} } _{e} (t)  \text{  \ \ \  \ \ } \hat{\kappa} _{e}  (t)   \\  
\pm \hat{\kappa} _{e} ^{\ast}  (t) \text{ \ \ } -\hat{ {\cal H} } _{e} ^{ \ast } (t) 
\end{array}
\right],
\end{eqnarray}
where $\kappa _{e} (t)$ represents the pairing field and $\hat{ {\cal H} } _{e} (t) = \hat{H} _{e} (t) - \mu $ with $\hat{H} _{e} (t)$ and $\mu$ as the  Hamiltonian and chemical potential for electrons. The two signs in $\pm$ have been introduced for different BCS models. The effective density matrix is \cite{Valatin}
\begin{eqnarray}
\hat{ \rho } _{b}(t) = \left[
\begin{array}{c}
\hat{\rho} _{e} (t) \text{ \ \ \ \ \ \  }  \hat{\Delta} _{e}(t) \\
\text{ \ } \pm \hat{\Delta} _{e} ^{\ast} (t) \text{ \ \ \ } \hat{I}_{e}-\hat{\rho} _{e} ^{ \ast } (t)
\end{array}
\right],
\end{eqnarray}
where $\hat{\rho} _{e} (t)$ denotes the one-body density matrix for electrons, $\hat{\Delta} _{e} (t)$ is the pairing tensor, and $\hat{I} _{e}=\hat{I} _{e} ^{\ast}$ is the identity operator for electron wavefunctions. The operators $\hat{H} _{b}(t)$ and $\hat{\rho} _{b} (t)$ follow
\begin{eqnarray}
\hat{S}_{b} \hat{\rho} _{b} ^{\ast} (t) \hat{S} _{b} ^{ \dagger } = \hat{I} _{b} - \hat{\rho} _{b}(t) 
\end{eqnarray}
\begin{eqnarray}
\hat{S} _{b} \hat{H} _{b} ^{ \ast } (t) \hat{S} _{b} ^{ \dagger } = - \hat{H} _{b} (t).
\end{eqnarray}
Here 
\[
\hat{I} _{b}= \left[ \begin{array}{c} \hat{I}_{e} \text{ \ } 0 \\ \text{ } 0 \text{ \ } \hat{I} _{e} \end{array} \right]
\]
 is the identity operator for Bogoliubov-BCS quasiparticles, and 
\[
 \hat{S} _{b} \equiv \left[ \begin{array}{c} \text{  }0 \text{ \ \ } \hat{I} _{e} \\ \mp \hat{I} _{e} \text{ \ }  0 \end{array} \right].
\]
We can see from the direct calculations that the Hamiltonian and density matrix satisfying Eqs. (10) and (11) can be wriiten as the forms given by  Eqs. (8) and (9). To model Bogoliubov-BCS quasiparticles by Eq. (1), we shall take $\hat{\rho} _{b} (t)$ and $\hat{H} _{b} (t)$ as the density matrix and Hamiltonian in such an euqation. Let $ \hat{\Gamma}_{b} (t)$ and $ \hat{\Gamma} _{b} ^{\prime} (t)$ as $\hat{\Gamma} (t)$ and $\hat{\Gamma} ^{\prime}(t)$ for the lifetimes of quasiparticles and quasiholes. It will be shown in this section that we shall introduce the constraint 
\begin{eqnarray}
\hat{S} _{b} \hat{ \Gamma }_{b}  ^{\ast} (t) \hat{S} _{b} ^{ \dagger }= \hat{\Gamma} ^{\prime} _{b}(t)
\end{eqnarray}
to preserve Eq. (9). The lifetimes of Bogoliubov-BCS quasiparticles and quasiholes are related under the above equation, which is reasonable because we can obtain the density matrix for quasiholes from that for quasiparticles by Eq. (10)

To see why we need to introduce the above constraint, first we consider a s-wave pairing case where the Bogoliubov transformation \cite{Bogoliubov,Entin-Wohlman} 
\begin{eqnarray}
\gamma _{ {\bf k} \uparrow } = z _{{\bf k}} c _{ {\bf k} \uparrow } + z ^{\prime} _{{\bf k}} c ^{\dagger} _{-{\bf k} \downarrow } 
\end{eqnarray}
\[
\gamma _{ -{\bf k} \downarrow } = - z _{{\bf k}} ^{\prime} c ^{\dagger} _{ {\bf k} \uparrow } + z _{{\bf k}} c _{ -{\bf k} \downarrow } 
\]
yields the annihilators for the quasiparticle with the excited energy $\xi _{{\bf k}} >0$ around a specific time $t_{1}$. Here the complex numbers $z _{{\bf k}}$ and $z ^{\prime} _{{\bf k}}$ denote the transformation coefficients. Let $ {\bf u} _{b} ({\bf r};{\bf k}+)$ and ${\bf u} _{b} ({\bf r};{\bf k}-) \equiv \hat{S} _{b} {\bf u} _{b} ^{\ast} ({\bf r};{\bf k}+)$ be the positive- and negative-energy orbitals corresponding to the quasiparticles annihilated by $ \gamma _{ {\bf k} \sigma }$, and assume that the Hamiltonian can be approximated as $\hat{H} _{b} = \sum _{{\bf k}}  \xi _{{\bf k}} [ {\bf u} _{b} ({\bf r};{\bf k}+) {\bf u} _{b} ^{\dagger} ({\bf r} ^{\prime};{\bf k}+)- {\bf u} _{b} ({\bf r};{\bf k}-) {\bf u} _{b} ^{\dagger} ({\bf r} ^{\prime};{\bf k}-)]$, the unperturbed Hamiltonian, as $t \sim t_{1}$. The number of quasiparticles in orbital ${\bf k}$ is $F _{b} ({\bf k};t) = tr [ \hat{\rho}_{b} (t) {\bf u} _{b} ({\bf r};{\bf k}+) {\bf u} _{b} ^{\dagger} ({\bf r } ^{\prime}; {\bf k}+)]$. We also have $F _{b} ({\bf k};t)=1-tr [ \hat{\rho}_{b} (t) {\bf u} _{b}({\bf r};{\bf k}-) {\bf u} _{b} ^{\dagger} ({\bf r};{\bf k}-)]$ because
\begin{eqnarray}
tr [\hat{\rho}_{b} (t) {\bf u} _{b} ({\bf r};{\bf k}+) {\bf u} _{b} ^{\dagger} ({\bf r } ^{\prime};{\bf k}+) ] = 1-tr [ \hat{\rho}_{b} (t) {\bf u} _{b}({\bf r};{\bf k}-) {\bf u} _{b} ^{\dagger} ({\bf r};{\bf k}-)]
\end{eqnarray}
from Eq. (10). Let $2 \alpha _{b} ( {\bf k},t )$ and $2 \beta _{b} ( {\bf k    },t )$ be the loss and gain rates of the quasiparticles in orbital ${\bf k}$ when $t \sim t _{1}$. The loss (gain) in orbital ${\bf k}$ not only decreases (increases) the occupation number in $ {\bf u} _{b} ({\bf r};{\bf k}+)$, but also increases (decreases) the occupation number in $ {\bf u} _{b} ({\bf r};{\bf k}-)$ because of the above equation. By checking the loss and gain rates, we shall set
\begin{eqnarray}
\hat{\Gamma} _{b} (t)= \sum _{{\bf k}} \alpha _{b} ({\bf k};t) {\bf u} _{b} ({\bf r};{\bf k}+){\bf u} ^{\dagger} _{b} ({\bf r}^{\prime};{\bf k}+) + \hat{\Gamma} _{b} ^{(l-)} (t)
\end{eqnarray}
\begin{eqnarray}
\hat{\Gamma} ^{\prime} _{b} (t)= \sum _{{\bf k}} \beta _{b} ({\bf k};t) {\bf u} _{b} ({\bf r};{\bf k}+){\bf u} ^{\dagger} 
_{b} ({\bf r}^{\prime};{\bf k}+) + \hat{\Gamma} _{b} ^{(g-)} (t)
\end{eqnarray}
with the two operators
\begin{eqnarray}
\Gamma _{b} ^{(l-)} = \hat{S} _{b} \left( \sum _{{\bf k}} \beta _{b} ({\bf k};t) {\bf u} _{b} ({\bf r};{\bf k}+){\bf u} ^{\dagger} _{b} ({\bf r}^{\prime};{\bf k}+) \right) ^{\ast}  \hat{S} _{b} ^{\dagger}= \sum _{{\bf k}} \beta _{b} ({\bf k},t) {\bf u} _{b} ({\bf r};{\bf k}-){\bf u} ^{\dagger} _{b} ({\bf r}^{\prime};{\bf k}-)
\end{eqnarray}
\begin{eqnarray}
 \Gamma _{b} ^{(g-)} = \hat{S} _{b} \left( \sum _{{\bf k}} \alpha _{b} ({\bf k},t) {\bf u} _{b} ({\bf r};{\bf k}+){\bf u} ^{\dagger} _{b} ({\bf r}^{\prime};{\bf k}+) \right) ^{\ast} \hat{S} _{b} ^{\dagger} = \sum _{{\bf k}} \alpha _{b} ({\bf k},t) {\bf u} _{b} ({\bf r};{\bf k}-){\bf u} ^{\dagger} _{b} ({\bf r}^{\prime};{\bf k}-).
\end{eqnarray}
In comparison with Eq. (4), the first terms at the right hands of Eqs. (15) and (16) can be taken as the extensions of $\hat{\Gamma} _{e}$ and $\hat{\Gamma} ^{\prime} _{e}$ to BCS models by $u _{e} ({\bf r};{\bf k}) \rightarrow {\bf u} _{b} ({\bf r};{\bf k}+)$, $\alpha _{e} ({\bf k};t) \rightarrow \alpha _{b} ({\bf k};t)$, and $\beta _{e} ({\bf k};t) \rightarrow \beta _{b} ({\bf k};t)$. On the other hand, $\hat{\Gamma} _{b} ^{(l-)} $ and $ \hat{\Gamma} _{b} ^{(g-)} $ are to relate the occupation numbers in negative- and positive-energy orbitals. We can see that Eqs. (15) and (16) yield operators $\hat{\Gamma} _{b} (t)$ and $\hat{\Gamma} _{b} ^{\prime} (t)$ following Eq. (12) by direct calculations.

It is known that pairing creations/annihilations \cite{Bardeen1} should be incorporated in addition to quasiparticle transitions. By checking the loss and gain rates, we can obtain the coefficients 
\begin{eqnarray}
\alpha _{b} ({\bf k};t) = \frac{1}{2} \sum _{ {\bf k} ^{\prime} } \omega _{ {\bf k} ^{\prime} {\bf k} } ^{(T)} ( 1- F _{b} ( {\bf k}^{\prime} ;t  ) ) + \frac{1}{2} \sum _{ {\bf k} ^{\prime} } \omega _{ {\bf k} {\bf k} ^{\prime}} ^{(A)} F _{b} ( {\bf k} ^{\prime};t )   
\end{eqnarray}
\begin{eqnarray}
\beta _{b} ({\bf k};t) = \frac{1}{2} \sum _{ {\bf k} ^{\prime} } \omega _{ {\bf k} {\bf k} ^{\prime} } ^{(T)} F  _{b} ( {\bf k} ^{\prime};t ) + \frac{1}{2} \sum _{ {\bf k} ^{\prime} } \omega _{ {\bf k} {\bf k}^{\prime}} ^{(C)}( 1 - F _{b} ( {\bf k} ^{\prime} ;t ) )  
\end{eqnarray}
for the relaxation term due to the electron-phonon interaction. Here $\omega _{ {\bf k} {\bf k} ^{\prime} } ^{(T)}$, $\omega _{ {\bf k} {\bf k} ^{\prime} } ^{(A)}$, and $\omega _{ {\bf k} {\bf k} ^{\prime} } ^{(C)}$ represent positive real parameters for the quasiparticle transition from ${\bf k} ^{\prime}$ to ${\bf k}$, pairing annihilation for quasiparticles in ${\bf k}$ and ${\bf k} ^{\prime}$, and pairing creation for quasiparticles in ${\bf k}$ and ${\bf k} ^{\prime}$, respectively. Under Eqs. (15)-(20), the quantum master equation can be reduced as 
\begin{eqnarray}
\frac{\partial}{\partial t} F _{b} ( {\bf k},t ) = - 2 \alpha _{b} ( {\bf k} ;t) F _{b} ( {\bf k},t ) + 2 \beta _{b} ( {\bf k} ;t) (1- F _{b} ( {\bf k},t )) 
\end{eqnarray}
\[
= - \sum _{ {\bf k} ^{\prime} } \omega _{ {\bf k} ^{\prime} {\bf k} } ^{(T)} ( 1- F _{b} ( {\bf k}^{\prime} ;t  )  ) F _{b} ( {\bf k},t ) - \sum _{ {\bf k} ^{\prime} } \omega _{ {\bf k} {\bf k} ^{\prime}} ^{(A)} F _{b} ( {\bf k} ^{\prime};t ) F _{b} ( {\bf k},t )
\]
\[
+\sum _{ {\bf k} ^{\prime} } \omega _{ {\bf k} {\bf k} ^{\prime} } ^{(T)} F  _{b} ( {\bf k} ^{\prime} ;t ) (1- F _{b} ( {\bf k},t )) +  \sum _{ {\bf k} ^{\prime} } \omega _{ {\bf k} {\bf k}^{\prime}} ^{(C)}( 1 - F _{b} ( {\bf k} ^{\prime} ;t ) )(1- F _{b} ( {\bf k},t )).
\]
The above equation, in fact, is just the semiclassical master equation for Bogoliubov-BCS quasiparticles when the relaxation term is due to the electron-phonon interaction. \cite{Aronov}

In general, we can include the spin orientation and extend Eq. (14) as
\begin{eqnarray}
tr (\hat{\rho} _{b} (t) {\bf f} _{b} {\bf f} ^{\dagger} _{b} ) =  1 - tr [ \hat{\rho} _{b} (t)  ( \hat{S} _{b} {\bf f} ^{\ast}_{b} ) ( \hat{S} _{b} {\bf f} _{b} ^{\ast} ) ^{\dagger} ] 
\end{eqnarray}
for any normalized quasiparticle wavefunction ${\bf f}_{b}$. If $\hat{\Gamma} _{b} (t)$ ($\hat{\Gamma} _{b} ^{\prime}(t)$) induces the decrease (increase) of the number of quasiparticles in ${\bf f}_{b}$, $\hat{\Gamma} _{b} ^{\prime}(t)$ ($\hat{\Gamma} _{b} (t)$) should induce the increase (decrease) of the occupation number in $\hat{S} _{b} {\bf f} ^{\ast}_{b}$ based on the above equation. Therefore, we shall use Eq. (12) to relate $\hat{\Gamma} _{b} (t)$ and $\hat{\Gamma} _{b} ^{\prime}(t)$. When Eq. (12) is valid, the form given by Eq. (9) is preserved under Eq. (8) if the time evolution of the density matrix follows Eq. (1). To see this, assume that a fermionic density matrix $\hat{\rho} _{b} (t)$ follows Eq. (1) with $\hat{H} (t)= \hat{H} _{b} (t)$, $\hat{I}= \hat{I} _{b}$, $\hat{\Gamma} (t) = \hat{\Gamma} _{b} (t)$, and $\hat{\Gamma} ^{\prime} (t) = \hat{\Gamma} _{b} ^{\prime} (t)$. (Here an operator $\hat{O}$ is fermionic iff the inner product $\langle \alpha | \hat{O} | \alpha \rangle $ for any normalized ket $\alpha$ is a real number between 0 and 1.) In addition, assume that $\hat{S}_{b} \hat{\rho} _{b} ^{\ast} (t _{i}) \hat{S} _{b} ^{ \dagger } = \hat{I} _{b} - \hat{\rho} _{b}(t _{i})$ such that Eq. (10) holds true at the initial time  $t_{i}$, and let $\hat{\rho} ^{\prime} _{b} (t) \equiv \hat{S} _{b} (\hat{I} _{b}- \hat{\rho} _{b} ^{\ast} (t) ) \hat{S} _{b} ^{\dagger} = \hat{I} _{b} - \hat{S} _{b} \hat{\rho} _{b} ^{\ast} (t) \hat{S} _{b} ^{\dagger}$ when $t \geq t _{i}$. It is easy to see that $\hat{\rho} ^{\prime} _{b} (t _{i}) = \hat{\rho} _{b} (t _{i})$, and $\hat{\rho} ^{\prime} _{b} (t)$ is also fermionic. The time derivative of $\hat{\rho} ^{\prime} _{b} (t)$ follows
\begin{eqnarray}
\frac{ \partial }{ \partial t } \hat{\rho} ^{\prime} _{b} (t) = i \hat{S} _{b} [ \hat{\rho} _{b} ^{\ast} (t), \hat{H} _{b} ^{\ast} (t) ] \hat{S} _{b} ^{\dagger} + \hat{S}_{b} \{ \hat{\rho} _{b} ^{\ast} (t) , \hat{\Gamma} _{b} ^{\ast} (t) \} \hat{S} _{b} ^{\dagger} 
- \hat{S}_{b} \{ \hat{I} _{b} - \hat{\rho} _{b} ^{\ast} (t) , \hat{ \Gamma } _{b} ^{\prime \ast} (t) \} \hat{S} _{b} ^{\dagger}
\end{eqnarray}
\[
=i [ \hat{\rho} ^{\prime} _{b} (t), \hat{H} _{b} (t) ] + \{ \hat{I} _{b} - \hat{\rho} ^{\prime} _{b} (t) , \hat{\Gamma} ^{\prime} _{b} (t) \} - \{ \hat{\rho} ^{\prime} _{b} (t) , \hat{ \Gamma } _{b}(t) \} \text{ \ \ \ \ \ \ \ \ \  \  \ }
\]
from Eqs. (11) and (12). Hence $\hat{\rho} ^{\prime} _{b} (t)$ is also a fermionic density matrix following Eq. (1) with the initial condition the same as that for $\hat{\rho} _{b} (t)$. Because the uniqueness of the solution to Eq. (1) is expected under suitable assumptions, \cite{Huang2} we have $\hat{\rho} _{b} ( t ) = \hat{\rho} ^{\prime} _{b} (t) = \hat{S} _{b} (\hat{I} _{b}- \hat{\rho} _{b} ^{\ast} (t) ) \hat{S} _{b} ^{\dagger}$ for any time $t \geq t _{0}$. The equality implies Eq. (10), under which $\hat{\rho} _{b} (t)$ is of the form given by Eq. (9). Therefore, the form of $\hat{\rho} _{b} (t)$ is preserved under Eqs. (1) and (8) when $\hat{\Gamma} _{b} (t)$ and $\hat{\Gamma} _{b} ^{\prime} (t)$ satisfy Eq. (12). 

\section{The master equation for the coexistence of different order parameters}

In addition to superconducting electrons, other identical fermions may also form Bogoliubov-BCS quasiparticles by pairing. Equation (9) can be used for different Fermi systems if we replace $\hat{\rho} _{e} (t)$ and $\hat{\Delta} _{e} (t)$ by the density matrix and pairing tensor for the corresponding fermions. For fermionic Bogoliubov-BCS quasiparticles, we can substitute $\hat{\rho} _{b} (t)$ for $\hat{\rho} _{e} (t)$ at the right hand side of Eq. (9) to extend such type of quasiparticles by introducing the density matrix 
\begin{eqnarray}
\hat{ \rho } _{B}(t) = \left[
\begin{array}{c}
\hat{\rho} _{b} (t) \text{ \ \ \ \ \ \  }  \hat{\Delta} _{b} (t) \\
\text{ \ } \pm \hat{\Delta} _{b} ^{\ast} (t) \text{ \ \ \ } \hat{I}_{b}-\hat{\rho} _{b} ^{ \ast } (t)
\end{array}
\right].
\end{eqnarray}
Here $\hat{\Delta}_{b} (t)$ is the pairing tensor for the quasiparticles described by $\hat{\rho} _{b} (t)$, and $\hat{ \rho } _{B}(t)$ represents the extended Bogoliubov-BCS quasiparticles. Because $\hat{\rho} _{b} (t)$ and $\hat{I}_{b}-\hat{\rho} _{b} ^{ \ast } (t)$ are $2 \times 2$ matrices, $\hat{\Delta}_{b} (t)$ is also a $2 \times 2$ matrix and $\hat{ \rho } _{B}(t)$ is a $4 \times 4$ matrix. It will be shown in this section that different order parameters can be incorporated by introducing $\hat{ \rho } _{B}(t)$, and the following two constraints
\begin{eqnarray}
\hat{S} _{ B } \hat{\rho}_{ B } ^{\ast} (t) \hat{S} _{ B } ^{\dagger} = \hat{I} _{ B } - \hat{\rho}_{ B } (t)
\end{eqnarray}
\begin{eqnarray}
\hat{A} _{ B } \hat{\rho}_{ B } (t) \hat{A} _{ B } ^{\dagger} = \hat{\rho}_{ B } (t)
\end{eqnarray}
for $\hat{ \rho } _{ B }(t)$ can be introduced based on Eq. (10). Here the $4 \times 4$ matrices $\hat{S} _{ B } = \left[ \begin{array}{c} \hat{S}_{b} \text{ \ } 0 \\ \text{ } 0 \text{ \ \ } \hat{S} _{b} \end{array} \right]$, $\hat{I} _{ B }= \left[ \begin{array}{c} \hat{I}_{b} \text{ \ } 0 \\ \text{ } 0 \text{ \ \ } \hat{I} _{b} \end{array} \right]$, and $\hat{A} _{ B }=\left[ \begin{array}{c} \text{ \ \ } 0 \text{ \ \ \ } \hat{S}_{b} \\ \text{ } \mp \hat{S}_{b} \text{ \ } 0 \end{array} \right]$. The operator $\hat{I} _{ B }$ is just the identity matrix for the particles described by $\hat{ \rho } _{B}(t)$. To model $\hat{\rho} _{B} (t)$ by Eq. (1), let $\hat{ H } _{B}(t)$, $\hat{ \Gamma } _{B}(t)$, and $\hat{ \Gamma } ^{\prime} _{B}(t)$ as the corresponding $\hat{H} (t)$, $\hat{ \Gamma } (t)$, and $\hat{ \Gamma } ^{\prime} (t)$ in such an equation. It will be also shown in this section that we shall introduce
\begin{eqnarray}
\hat{S} _{ B } \hat{H} _{ B } ^{ \ast } (t) \hat{S} _{ B } ^{ \dagger } = - \hat{H} _{ B } (t)
\end{eqnarray}
\begin{eqnarray}
\hat{S} _{ B } \hat{ \Gamma }_{ B } ^{\ast} (t) \hat{S} _{ B } ^{ \dagger }= \hat{\Gamma} ^{\prime} _{ B }(t)
\end{eqnarray}
for the validity of Eq. (25), and introduce
\begin{eqnarray}
\hat{A} _{ B } \hat{H} _{ B } (t) \hat{A} _{ B } ^{ \dagger } = \hat{H} _{ B } (t),
\end{eqnarray}
\begin{eqnarray}
\hat{A} _{ B } \hat{ \Gamma }_{ B }  (t) \hat{A} _{ B } ^{ \dagger }= \hat{\Gamma} _{ B }(t)
\end{eqnarray}  
for Eq. (26). 

Because Eq. (24) is obtained from Eq. (9) by substituting $\hat{\rho}_{b}(t)$ for $\hat{\rho}_{e} (t)$, we can see that the constraint 
\begin{eqnarray}
\hat{S} _{ B } ^{\prime} \hat{\rho}_{ B } ^{\ast} (t) \hat{S} _{ B } ^{\prime \dagger} = \hat{I} _{ B } - \hat{\rho}_{ B } (t)
\end{eqnarray} 
with the $4 \times 4$ matrix $\hat{S} _{ B } ^{\prime}=\left[ \begin{array}{c} \text{ \ } 0 \text{ \ \ \ } \hat{I} _{b} \\ \text{ } \mp \hat{I} _{b} \text{ \ } 0 \end{array} \right]$ is equivalent to Eq. (24) from the equivalence between Eqs. (9) and (10). In addition, the following two constraints    
\begin{eqnarray}
\hat{S} ^{\prime} _{ B } \hat{H} _{ B } ^{ \ast } (t) \hat{S} _{ B } ^{ \prime \dagger } = - \hat{H} _{ B } (t)
\end{eqnarray}
\begin{eqnarray}
\hat{S} _{ B } ^{\prime} \hat{ \Gamma }_{ B }  ^{\ast} (t) \hat{S} _{ B } ^{ \prime \dagger }= \hat{\Gamma} ^{\prime} _{ B }(t)
\end{eqnarray}
should be introduced for Eq. (31) just as Eqs. (11) and (12) are introduced for Eq. (10). To use $\hat{\rho} _{b}(t)$ to represent the quasiparticles resulting from electron pairing, however, we shall introduce additional constraint for $\hat{\rho} _{b} (t)$ to follow Eq. (9).

We can see from the last section that Eq. (9) is preserved iff Eq. (10) holds true, and Eqs. (11) and (12) are important to the validity of Eq. (10). The $4 \times 4$ matrix $\hat{S} _{ B }=\left[ \begin{array}{c} \hat{S} _{ b } \text{ \ }  0 \\ 0 \text{ \ } \hat{S} _{ b } \end{array} \right]$ is the natural correspondence to the $2 \times\ 2$ matrix $\hat{S} _{ b }$, so it is reasonable to extend Eqs. (10)-(12) as Eqs. (25), (27), and (28). From Eq. (25), $\hat{\Delta} _{b} (t)$ is of the form 
\begin{eqnarray}
\hat{\Delta} _{b} (t) = \left[
\begin{array}{c}
\hat{\Delta} _{s} ^{\prime} (t) \text{ \ \ \ } \hat{\delta} _{s} (t) \\ \text{ }\pm \hat{\delta} ^{\ast} _{s} (t) \text{ \ } -\hat{\Delta} _{s} ^{\prime \ast} (t)
\end{array}
\right].
\end{eqnarray}

The matrix $\hat{A} _{ B }$ equals the product $\hat{S} _{ B } \hat{S} ^{\prime} _{ B }$. Therefore,
\begin{eqnarray}
\hat{A} _{ B } \hat{\rho} _{ B } (t) \hat{A} _{ B } ^{\dagger} = \hat{S} _{ B } ( \hat{S} ^{\prime} _{ B } \hat{\rho} _{ B } ^{\ast} (t) \hat{S} ^{\prime \dagger} _{ B } ) \hat{S} _{ B } ^{\dagger} = \hat{\rho}_{ B } (t)
\end{eqnarray} 
and we can obtain Eq. (26) from Eqs. (25) and (31). In fact, Eq. (26) is equivalent to Eq. (31) under Eq. (25), so we only need to consider Eqs. (25) and (26) for the form of $\hat{\rho} _{ B }(t)$ and $\hat{\rho} _{b} (t)$. By checking $\hat{A} _{ B } \hat{H}_{ B } (t) \hat{A} _{ B } ^{\dagger}$ and $\hat{A} _{ B } \hat{\Gamma}_{ B } (t) \hat{A} _{ B } ^{\dagger}$, we can see that Eqs. (29) and (30) are equivalent to Eqs. (32) and (33), respectively, if Eqs. (27) and (28) hold true. So we just need to take Eqs. (27)-(30) as the constraints on the quantum master equation. By considering the time-derivative of $\hat{S} _{B} (\hat{I} _{B}- \hat{\rho} _{B} ^{\ast} (t) ) \hat{S} _{B} ^{\dagger}$ and $\hat{S} _{B} ^{\prime} ( \hat{I} _{B}- \hat{\rho} _{B} ^{\ast} (t) ) \hat{S} _{B} ^{\prime \dagger}$, we can prove that the form of $\hat{\rho} _{B} (t)$ is preserved under these constraints.

From Eqs. (9), (24), and (34), we can rewrite $\hat{\rho} _{ B } (t)$ by 
\begin{eqnarray}
\hat{\rho} _{ B } (t) = \left[
\begin{array}{c}
\hat{\rho} _{e} (t) \text{ \ \ \ \ \ } \hat{\Delta} _{e} (t) \text{ \ \ \ \ \ } \hat{\Delta} ^{\prime} _{e} (t) \text{ \ \ \ \ \ \ } \hat{\delta} _{e} (t) \\
\pm \hat{\Delta} _{e} ^{\ast} (t) \text{ \ \ } \hat{I}_{e} - \hat{\rho} _{e} ^{\ast} (t) \text{ \ } \pm \hat{\delta} _{e} ^{\ast} (t) \text{ \ } -\hat{\Delta} _{e} ^{\prime \ast} (t) \\
\pm \hat{\Delta} _{e} ^{\prime \ast} (t) \text{ \ } \pm\hat{\delta} ^{\ast} _{e} (t) \text{ \ \ } \hat{I}_{e} - \hat{\rho} _{e} ^{\ast} (t) \text{ } -\hat{\Delta} _{e} ^{\ast} (t) \\
\hat{\delta} _{e} (t) \text{ \ \ \ } \mp \hat{\Delta} ^{\prime} _{e} (t) \text{ \ \ \ } \mp \hat{\Delta} _{e} (t) \text{ \ \ \ } \hat{\rho} _{e} (t)
\end{array}
\right]
\end{eqnarray}
Assume that ${\bf f} _{ B }$ be a normalized eigenfunction of $\hat{\rho} _{ B } (t)$ at a specific time $t=t _{1}$. Because $\hat{\rho} _{ B } (t _{1})$ is a $4 \times 4$ matrix, ${\bf f} _{ B }$ includes four component and we can write 
\begin{eqnarray}
{\bf f} _{ B }=\left( 
\begin{array}{c} 
f_{e1} \\ f_{h1} \\ f_{h2} \\ f_{e2} 
\end{array} 
\right).
\end{eqnarray} 
By checking the contribution of ${\bf f}_{ B } {\bf f} _{ B } ^{\dagger}$ to $\hat{\rho} _{ B } (t _{1})$, we can see that the components $f_{e1}f_{e1} ^{\ast}$ and $f_{e2}f_{e2} ^{\ast} $ are incorporated in the first and fourth diagonal terms, both of which are just $\hat{\rho} _{e} (t_{1})$. On the other hand, $f_{h1}f_{h1} ^{\ast}$ and $f_{h2}f_{h2} ^{\ast} $ are incorporated in the second and third diagonal terms, both of which are $\hat{I}_{e} - \hat{\rho} _{s} ^{\ast} (t)$. Because $\hat{\rho} _{e} (t_{1})$ and $\hat{I}_{e} - \hat{\rho} _{e} (t _{1})$ represent the density matrices for electrons and holes at $t _{1}$, we shall take $f _{e1}$ and $f _{e2}$ as electron components and take $f _{h1}$ and $f _{h2}$ as hole components. Therefore, ${\bf f}_{ B }$ contain two electron components and two hole components just as the four-component wavefunctions introduced for the coexistence of the superconducting and antiferromagnetic orders in Ref. \cite{Laughlin}. The coexistence may reveal the key to understand the high-temperature superconductors. The operator $\hat{\delta} _{e} (t)$ in Eq. (36) can correspond to the antiferromagnetic order while $\Delta ^{\prime} _{e}(t)$ and $\Delta _{e} (t)$ can be taken as the superconducting orders, and the extended master equation for $\hat{\rho} _{B} (t)$ could be used to model the nonequilibrium phenomena when there are different order parameters.

For the BCS-type quasiparticles described by $\hat{\rho} _{ B } (t)$, the density matrix for the corresponding quasiholes is $\hat{I} _{ B } - \hat{\rho} _{ B } (t)$. Just as mentioned in Ref. \cite{Huang2}, we can substitute $\hat{\rho} _{ B  } (t)$ for $\hat{\rho} _{e} (t)$ at the right hand side of Eq. (9) for the further extension. A chain of density matrices $ \hat{ \rho }_{ n } (t)$, in fact, can be constructed by generalizing Eqs. (9) and (24) as \cite{Huang2} 
\begin{eqnarray}
\hat{ \rho } _{ n+1 }(t) \equiv \left[
\begin{array}{c}
\hat{\rho} _{n} (t) \text{ \ \ \ \ \ \  }  \hat{\Delta} _{n} (t) \\
\text{ \ } \pm \hat{\Delta} _{ n } ^{\ast} (t) \text{ \ \ \ } \hat{I}_{ n }-\hat{\rho} _{ n } ^{ \ast } (t)
 \end{array} \right]
\end{eqnarray}
for any positive integer $n$ if we set $\hat{ \rho }_{ 1 }(t) \equiv \hat{ \rho} _{e} (t)$ and $\hat{ I }_{ 1 }(t) \equiv \hat{ I} _{e} (t)$. Here $\hat{\Delta} _{n} (t)$ and $\hat{I} _{n}$ denote the corresponding pairing fields and identity operators. (Under the above equation, $ \hat{\rho} _{b} (t) = \hat{\rho} _{2} (t) $, $ \hat{I} _{b} (t) = \hat{I} _{2} (t) $, $ \hat{\rho} _{B} (t) = \hat{\rho} _{3} (t) $, and $ \hat{I} _{B} (t) = \hat{ I } _{3} (t) $ .) On the other hand, we can extend the Bogoliubov-BCS density matrix by considering the quasiparticles with multiple electron components and the coupling between them and the corresponding quasiholes. \cite{Huang4} The orbial ${\bf f} _{ B }$ in Eq. (37) can be interpreted as coupling between the antiferromagnetic-like quasiparticles
\begin{eqnarray}
\left(\begin{array}{c} f_{e1} \\ f_{e2} \end{array} \right)
\end{eqnarray}
and the corresponding quasiholes 
\begin{eqnarray}
\left(\begin{array}{c} f_{h1} \\ f_{h2} \end{array} \right).
\end{eqnarray}
There is no upper limit to the number of components in principal, and it may be convenient to introduce the extra dimensions for the quasiparticles with the infinite components. 

\section{Discussion}

The corresponding Hamiltonian ${\hat{H} _{B} (t)}$ in the last section, in fact, is of the form
\begin{eqnarray}
\hat{H} _{ B } (t) = \left[
\begin{array}{c}
\hat{ {\cal H} } _{e} (t) \text{ \ \ \ \ \ } \hat{\kappa} _{e} (t) \text{ \ \ \ \ \ } \hat{\kappa} ^{\prime} _{e} (t) \text{ \ \ \ \ \ \ } \hat{ v } _{e} (t) \\
\pm \hat{\kappa} _{e} ^{\ast} (t) \text{ \  } - \hat{ {\cal H} } _{e} ^{\ast} (t) \text{ \ } \pm \hat{ v } _{e} ^{\ast} (t) \text{ \ } -\hat{\kappa} _{e} ^{\prime \ast} (t) \\
\pm \hat{\kappa} _{e} ^{\prime \ast} (t) \text{  \ } \pm\hat{ v } ^{\ast} _{e} (t) \text{  \  } - \hat{ {\cal H} } _{e} ^{\ast} (t) \text{ \  } -\hat{\kappa} _{e} ^{\ast} (t) \\
\hat{ v } _{e} (t) \text{ \ \ \  } \mp \hat{\kappa} ^{\prime} _{e} (t) \text{ \  } \mp \hat{\kappa} _{e} (t) \text{ \ \ \ } \hat{ {\cal H} } _{e} (t)
\end{array}
\right]
\end{eqnarray}
iff Eqs. (27) and (29) are valid. For the time-independent case, the eigenstate of the above Hamiltonian is of the form given by Eq. (37), in which $f _{e1}$ and $f_{e2}$ are the electron components and $f _{h1}$ and $f_{h2}$ are the two hole components. When $\hat{\kappa} _{s} = \hat{\kappa} ^{\prime} _{s}=0$, there is no coupling between the electron and hole components and $\hat{H} _{B}$ can be reduced as  
\begin{eqnarray}
\hat{H} _{a} = \left[
\begin{array}{c}
\hat{ {\cal H} } _{e} \text{ \ \ }  \hat{ {v} } _{e}   \\  \text{ } \hat{v} _{e}  \text{ \ } \hat{ {\cal H} } _{e} 
\end{array}
\right],
\end{eqnarray}
of which the eigenstate is of the form given by Eq. (39) and includes only two electron components just as the antiferromagnetic-quasiparticle eigenstate. In addition, Eq. (29) is reduced as 
\begin{eqnarray}
\hat{A} ^{\prime} \hat{H} _{a} \hat{A} ^{ \prime \dagger} = \hat{H} _{a}, 
\end{eqnarray}
where
\begin{eqnarray}
\hat{A} ^{\prime} =\left[ \begin{array}{c} 0 \text{ \ } \hat{I}_{e} \\ \text{ } \hat{I}_{e} \text{ \ } 0 \end{array} \right].  
\end{eqnarray}

For the $d _{ {x^{2}} \text{-} {y^{2}} }$ density wave (DDW) model \cite{Bena}, the wavevector ${\bf k}=(k _{x}, k _{y})$ satisfies $max( k _{x} , k _{y}) \leq \pi$. In addition to the noninteracting Hamiltonian given by Eq. (2), we shall include the antiferromagnetic term ${\cal W} = \sum _{ {\bf k} \in ABZ } W _{ {\bf k} } c _{ {\bf k}, \sigma } ^\dagger  c _{ {\bf k}+{\bf Q}, \sigma} + W _{ {\bf k} } ^{\ast}  c _{ {\bf k}+{\bf Q} , \sigma} ^\dagger  c _{ {\bf k}, \sigma}$.  Here ABZ denotes the antiferromagnetic Brillouin zone where $| {\bf k} _{x} | + | {\bf k} _{y} | \leq \pi$, the vector ${\bf Q}= (\pi, \pi) $, and each $W _{ {\bf k} }$ is a complex number. Let $\hat{ {\cal W} } = \sum _{ {\bf k} \in ABZ, \sigma }  W _{ {\bf k} } u_{e} ({\bf r};{\bf k}) u_{e} ^{\ast} ({\bf r};{\bf k} + {\bf Q}) + W _{ {\bf k} } ^{\ast} u_{e} ({\bf r};{\bf k}+ {\bf Q}) u_{e} ^{\ast} ({\bf r};{\bf k} )$ be the effective term corresponding to ${\cal W}$, the DDW effective Hamiltonian is
\begin{eqnarray}
\hat{H} _{DDW} =  \left[
\begin{array}{c} \text{ \ \ \ }\hat{P} _{ABZ} \hat{ H _{o}}  \hat{P} _{ABZ}  \text{  \ \ \ \ \ \ \ \ \ \ \  }   \hat{P} _{ABZ} \hat{ {\cal W} } (\hat{I} _{e} -  \hat{P} _{ABZ} ) \text{ \ \ \ \ \ }  \\  (\hat{I} _{e} -  \hat{P} _{ABZ} ) \hat{ {\cal W} } ^{\dagger}  \hat{P} _{ABZ}  \text{ \ \ \ \ \ } ( \hat{ I } _{e} - \hat{P} _{ABZ} )  \hat{ H _{o}}   ( \hat{ I } _{e} - \hat{P} _{ABZ} )   
\end{array}
\right] - \mu \hat{ I _{b} } .
\end{eqnarray}
Here the projection operator $\hat{P} _{ABZ}$ is to project any electron wavefunction into ABZ. Each eigenket of $\hat{H} _{DDW}$ is of the following form       
\begin{eqnarray}
{\bf u} _{a} ({\bf r};{\bf k}) =  Z_{ {\bf k} } u_{e} ({\bf r};{\bf k} ) \left(\begin{array}{c} 1 \\  0 \end{array} \right)  +  Z_{ {\bf k} } ^{\prime}  u_{e} ({\bf r};{\bf k} +{\bf Q}) \left(\begin{array}{c} 0 \\ 1 \end{array} \right),
\end{eqnarray}
where the coefficients $Z_{ {\bf k} } $ and $ Z_{ {\bf k} } ^{\prime}$ satisfy
\begin{eqnarray}
\left[
\begin{array}{c}
\varepsilon _{ {\bf k} } - \lambda _{ {\bf k} } \text{ \ \ \ \ \ \ \ \ }  W _{ {\bf k} }   \\  \text{ \ \ \ \ \ \ } W _{ {\bf k} } ^{\ast} \text{ \ \ \ \ \ \ } \varepsilon _{ {\bf k}+{\bf Q} } - \lambda _{ {\bf k} } 
\end{array} 
\right]
\left[
\begin{array}{c}
Z_{ {\bf k} } \\ Z_{ {\bf k} } ^{\prime}
\end{array} 
\right]=0
\end{eqnarray}
with $\lambda _{ {\bf _k} }$ as the eigenenergy. It is natural that 
\begin{eqnarray}
{\bf u} _{a} ^{\prime} ({\bf r};{\bf k}) =   Z_{ {\bf k} } u_{e} ({\bf r};{\bf k} ) \left(\begin{array}{c} 0 \\  1 \end{array} \right)  +  Z_{ {\bf k} } ^{\prime}  u_{e} ({\bf r};{\bf k} +{\bf Q}) \left(\begin{array}{c} 1 \\ 0 \end{array} \right) 
= \hat{A} ^{\prime} {\bf u} _{a} ({\bf r};{\bf k}) 
\end{eqnarray}
plays the same role (with the same eigenenergy $\lambda _{ {\bf _k} }$) as $ {\bf u} _{a} ({\bf r};{\bf k})$ since ${\bf u} _{a} ^{\prime} ({\bf r};{\bf k})$ can be obtained by using $\hat{A} ^{\prime}$ to exchange the first and second components of ${\bf u} _{a} ({\bf r};{\bf k})$. For the DDW model, actually we can construct the effective Hamiltonian $\hat{H} _{D} $
\begin{eqnarray}
\hat{H} _{D} = \hat{H} _{DDW} + \hat{A} ^{\prime} \hat{H} _{DDW} \hat{A} ^{\prime \dagger}
\end{eqnarray}
such that both Eqs. (46) and (48) provide the eigenkets. The Hamiltonian $\hat{H} _{D}$, in fact, is just a specific time-independent form of $\hat{H} _{a}$ because 
\begin{eqnarray}
\hat{A} ^{\prime} \hat{H} _{D} \hat{A} ^{ \prime \dagger} = \hat{H} _{D}.
\end{eqnarray}
For each $\lambda _{ {\bf k} }$, $\hat{H}_{D}$ has two degenerate eigenstates ${\bf u} _{a}$ and ${\bf u} _{a} ^{\prime}$. 

When $\hat{ H } _{a}$ is time-independent, we can diagaonalize it by the eigenkets of $\hat{A} ^{\prime}$ because of Eq. (43). The operator $\hat{A} ^{\prime}$ has only two eigenvalues $+1$ and $-1$, and the corresponding eigenkets are of the forms 
\begin{eqnarray}
\frac{f^{(1)} _{m} ({\bf r}) }{ \sqrt{2} } \left(
\begin{array}{c} 1 \\ 1 \end{array}  \right)
\text{ and }  
\frac{ f^{(2)} _{m} ({\bf r}) }{ \sqrt{2} } \left(
\begin{array}{c} 1 \\ -1 \end{array} 
\right),  
\end{eqnarray}
respectively. The functions $f^{(1)} _{m} ({\bf r})$ and $f^{(2)} _{m} ({\bf r})$ should satisfy 
\begin{eqnarray}
\hat{H} ^{(1)} _{e} f ^{(1)} _{m} = \varepsilon _{m} f ^{(1)} _{m} \text{ and } \hat{H} _{e} ^{(2)} f ^{(2)} _{m}  = \varepsilon _{m} f ^{(2)} _{m}
\end{eqnarray}
to determine the eigenvalues and eigenfunctions of $\hat{H} _{a}$, where 
\begin{eqnarray}
 \hat{H} ^{(1)} _{e} = \hat{H}_{e} + \hat{ v } _{e} \text{ and } \hat{H} ^{(2)} = \hat{H}_{e} - \hat{ v} _{e}.
\end{eqnarray}
In fact, we can take $\hat{{\cal H} }_{e}=( \hat{H} _{e} ^{(1)} + \hat{H} _{e} ^{(2)} )/2 - \mu \hat{ I } _{e} $  and $\hat{ v } _{e} =( \hat{H} ^{(1)} _{e} - \hat{H} ^ {(2)} _{e} )/2$ to construct $\hat{H} _{a}$ by Eq. (42) when two electron-like Hamiltonians $\hat{H} ^{(1)} _{e}$ and $ \hat{H} ^{(2)} _{e}$ are given. For a chemical bond, as shown in Appendix,we can include the correlation between covalent and ionic states by using two Hamiltonians $\hat H ^{(1)} _{e}$ and $\hat H^{(2)} _{e}$.

For each $\hat{\rho} _{n+1}$ with $n >1$ constructed by Eq. (38), we can  reduce the quantum master equation into $2 ^{n-1}$ irreversible equations for electron parts after decoupling the electron and hole components. We note that such decoupling yields Hamiltonians such as $\hat{H} _{a}$ similar to those introduced for fractal structures \cite{fractal}, and the single-particle scheme of Pershin et al. \cite{Pershin} are based on a set of irreversible equations.

\section{Conclusion}                                             

A quantum master equation is obtained for the density matrix representing Bogoliubov-BCS quasiparticles. Such an equation can be reduced to the semiclassical equation, and can be extended for the coexistence of different order parameters. 

\section*{Appendix}

Consider a chemical bond between two atoms X and Y (based on the linear combination of atomic orbitals), and assume that the electronegativity of atom Y is much higher than that of atom X. Therefore, the ionic state
\begin{eqnarray}
| \Psi _ {1}\rangle = c _{Y \uparrow} ^{\dagger} c _{Y \downarrow} ^{\dagger} |0 \rangle
\end{eqnarray} 
is dominated and we do not need to consider the probability for both electrons to occupy the orbital close to atom X. Here $| 0 \rangle$ denotes the vacuum state, and $c _{Y \uparrow} ^{\dagger}$ ($c _{Y \downarrow} ^{\dagger}$) is the creator to occupy the up-spin (down-spin) orbital near atom Y in such a chemical bond. To include the covalent contribution \cite{Hilberty}, we shall consider the covalent state 
\begin{eqnarray}
| \Psi _{2} \rangle = \frac{1}{\sqrt{2}} ( c _{X \uparrow} ^{\dagger} c _{Y \downarrow} ^{\dagger} + c _{Y \uparrow} ^{\dagger} c _{X \downarrow} ^{\dagger} )|0 \rangle,
\end{eqnarray}
where $ c _{X \uparrow} ^{\dagger}$ ($ c _{X \downarrow} ^{\dagger}$) is the creator following $\{ c _{X \sigma} , c _{Y \sigma} ^{\dagger} \} =0$ for the up-spin (down-spin) orbital dominated by the component belonging to atom X. Therefore, the ground-state wavefunction is
\begin{eqnarray}
| \Psi \rangle = C _{1} | \Psi _{1} \rangle + C_{2} | \Psi _{2} \rangle
\end{eqnarray}
with the coefficients $C _{1}$ and $C _{2}$ satisfy $ | C _{1} | ^{2} + | C _{2} | ^{2} =1$.

Let $\phi _{X} ( {\bf r} )$ ($\phi _{Y} ( {\bf r} )$) be the normalized spatial part for the up- and down-spin orbitals of atom X (Y) in such a chemical bond. Because one electron should be located at the orbital of atom Y in both $| \Psi _{1} \rangle $ and $| \Psi _{2} \rangle $, it is natural to include $\hat{ \rho } ^{(1)} ( {\bf r} , {\bf r} ^{\prime} ) = \phi _{Y} ( {\bf r} ) \phi _{Y} ^{\ast} ( {\bf r} ^{\prime} ) $ as the density matrix for one corresponding quasiparticle. On the other hand, we can set $\hat{ \rho } ^{(2)} ( {\bf r} , {\bf r} ^{\prime} )= \phi _{L} ( {\bf r} ) \phi _{L} ^{\ast} ( {\bf r} ^{\prime} )$ with $ \phi _{L} ( {\bf r} ) = C _{1} \phi _{Y} ( {\bf r} ) + C _{2} \phi _{X} ( {\bf r} ) $ as the density matrix for the other quasiparticle to include effects due to the linear combination of $| \Psi _{1} \rangle $ and $| \Psi _{2} \rangle $. The total energy ${\cal E}$, which is due to Coulomb potential $ U ({\bf r} _{1} , {\bf r} _{2}) = 1/4 \pi \varepsilon | {\bf r} _{1} - {\bf r} _{2} |$ in addition to the Hamiltonian $\hat{h}$ including the kinetic term and external field, for such a chemical bond is
\begin{eqnarray}
{\cal E} = tr [ \hat{h} (\hat{\rho} ^{(1)} + \hat{\rho} ^{(2)})] + \int d^{3} r_{1} d ^{3} r _{2} U  ({\bf r} _{1} , {\bf r} _{2}) \hat{\rho} ^{(1)}({\bf r}_{1},{\bf r}_{1}) \hat{\rho} ^{(2)}({\bf r}_{2},{\bf r}_{2})  
\end{eqnarray}
\[
 + (\sqrt{2} - 1) [  tr\hat{h}\hat{d} + \int d^{3} r_{1} d ^{3} r _{2}  U ({\bf r} _{1} , {\bf r} _{2}) \hat{\rho} ^{(1)}({\bf r}_{1},{\bf r}_{1}) \hat{d} ({\bf r}_{2},{\bf r}_{2}) ]  \text{ \ \  }
\]
\[
 + \int d^{3} r_{1} d ^{3} r _{2} U ({\bf r} _{1} , {\bf r} _{2}) \hat{\rho} ^{(1)}({\bf r}_{1},{\bf r}_{2}) \hat{d} ^{\prime} ({\bf r}_{2},{\bf r}_{1}). \text{ \ \  \ \ \ \ \ \ \ \ \ \ \ \ \ \ \ \ \ \ \ \ \ \ }
\]
Here $\varepsilon $ denotes the dielectric constant, $\hat{d} \equiv \hat{\rho} ^{(1)} \hat{\rho} ^{(2)} (\hat{I} _{e} - \hat{\rho} ^{(1)}) +  (\hat{I} _{e} - \hat{\rho} ^{ (1) })\hat{\rho} ^{ (2) } \hat{\rho} ^{ (1) }$, and $\hat{d} ^{\prime} \equiv (\hat{I} _{e} - \hat{\rho} ^{(1)}) \hat{\rho} ^{(2)} (\hat{I} _{e} - \hat{\rho} ^{(1)})$. We can consider the variation on the energy ${\cal E}$ to obtain the effective Hamiltonains $\hat{H} 
_{e} ^ {(1)} \equiv \delta {\cal E}/ \delta \hat{\rho} ^{(1)}  $ and $\hat{H} ^{(2)} _{e} \equiv \delta {\cal E}/ \delta 
\hat{\rho} ^{(2)} $.

\end{document}